# Strategy For Assessment Of Land And Complex Fields Type Analysis Through GIS In Bangladesh.

Yeasir Fathah Rumi[1], Uzzal Kumar Prodhan[2], Mohammed Ibrahim Hussain[3], A.H.M. Shahariar Parvez[3], Md. Ali Hossain[4]

[1,3] Assistant Professor, CSE Dept., Bangladesh University, Dhaka, Bangladesh
yeasirfathah@yahoo.com, ibrahim180772@gmail.com, sha0131@yahoo.com

[2] Assistant Professor, CSE Dept.,Jatiya Kabi Kazi Nazrul Islam University,Trishal, Mymensingh, Bangladesh
uzzal_bagerhat@yahoo.com

[4] Lecturer,CSE Dept., Bangladesh University, Dhaka, Bangladesh
ali.cse.bd@gmail.com

*ABSTRACT*

*Bangladesh is an over populated developing country where crisis of food is a major issue, it faces different infrastructure problem in every sector. For Poverty Alleviation from the country we have to confirm cultivable land to increase the crop production for feeding the over population of the country. This paper focuses on the measurement of cultivable land for cultivation. The main purpose of this paper is to briefly describe how the GIS, Digital Mapping, Internet concepts and tools can effectively contribute in the modeling, analysis and visualization phases within an engineering or research project according to the crops by using object detection, object tracking and field mapping in Bangladesh. Through GIS mapping of the agricultural lands, the statistics can be made of how much land is cultivable and each year how much land we are losing. Mapping the cultivation land will tell us how much crop we have to import from other countries. Enabling real-time GIS analysis anytime, anywhere, the implementation of the GIS information to a wider aspect. Automation is the indicator of the modern civilizations. The system will benefit the food stock of the country according to the harvest. For this research we developed a new interactive system. The system will integrate with GIS project data in Google Earth, first finds highly accurate cluster images and partial images, obtains user feedback to merge or correct these digests, and then the supplementary visual analysis complete the partitioning of the data. This study was conducted at the software laboratory, Computer Science and Engineering department, Jahangirnagar University, Dhaka, Bangladesh in 2013.*

## 1. INTRODUCTION

Geographical Information Systems (GIS) are such types of systems which are designed to store and control data relating to locations depending on the earth's surface. GIS takes and analyze different layers of information in the shape of maps and satellite images easily and allows identifying the spatial affairs. The use of GIS application has been introduced in Bangladesh for major impact in influencing policy makers for better use of information for planning and development, its exposure is limited to a small community of GIS users. The potential for GIS applications in Bangladesh is significant. GIS can be used as a tool for development and planning. Now a day's food export has decreased significantly in the world. Countries like China, India, and Vietnam has barred their food grain exporters to export food grains. In order to secure the food safety program, it is not very hard to assume that, in case of food crisis or deficit we have to face a acute crisis to feed our huge population. So it is very necessary to have proper data about arable land in total, its declining rate per year to take precautionary measures as well as detrimental consequences. Our information and telecommunication technology is going to cover up 86% of the world population through internet and mobile network and it is expected by the year of 2050 more than 80% of the world population will be using the internet. World information society ensures few things which will be achieved by the year of 2015. More than half of the world population having access to ICT by 2015 which will connect the villages together through community access points. Connecting schools,colleges, universities with ICT. Connecting engineering and research institutions with ICT, Connecting public libraries, cultural centers museums, post offices, hospitals and archives with ICT and upgrading all primary and secondary school curricula to meet the targets of information society [5]. This system will provide information to farmers about harvest and crops so that they can ensure along with appropriate information. Farmers will be advanced into storing crops and can provide increasing amounts of comparative information of their crops through well trained staff. As well as giving additional information to farmers. Analysis of GIS information is becoming a core competence of farmers in the move away to relationship business. ICT organizations can offer value added information in this perspective. The media, magazines and newspapers, television and radio programs all can provide latest updates on analysis. Internet user will always search on new crops. The search can link the farmers all together to take decision about what crops they will grow for the season. New technology is not a threat to the farmers, it is considerable opportunity to exploit and shape new targets. Experts could define basic parameters and areas ensure that they would integrate with existing systems then this will be the dominant technology. Due to the large geographical field type analysis the success of IT strategy depends on the ability to network all around the world.

## 2. METHODOLOGY

In order to compute the complexity of the GIS algorithm, the last query retrieves every that lies within 500 meters of one specific road. The query can be made more general by dropping the second part of the where clause and all towns that lie within meters of a road In a sense, this is the optimal type of integration of spatial analysis in a GIS However certain types of analysis are more difficult execution this way For example the query which ends the shortest path from A to B over the network of road to express in Post quell An alternative is to regard the road network as one complex object in which the individual  roads sub objects are linked Now a shortest path function can be applied to the complex object road network Unfortunately this solution is also not feasible in many DBMSs .In general It is more difficult to perform complex spatial analyses with one single database query if the operands are complex objects. A feasible way to implement this is by using separate dedicated programs [1]. Almost every day we find news of land grabbing, destruction of forest trees, removal of sand from river beds, filling up of ponds, destruction of crops cultivation lands, etc in our daily newspapers. From the ongoing situation it seems like our respective government institutions are lacking sufficient manpower to keep watch of the national resources or private resources to uphold the governmental environment protection related policies. Many private land boundary related disputes could be handled efficiently using GIS with accuracy and efficiency.

The use of GIS application in Bangladesh started in 1991 by ISPAN (Irrigation Support Project for Asia and the Near East) for the FAP-19 (Flood Action Plan-19) project . The organization is now named as EGIS (Environmental and GIS Support Projects for Water Sector Planning). At present there are over 50 GIS installations in the country. At the beginning, most of the GIS installations were donor supported and operated by foreign experts with limited local personnel. Now the situation has changed, a number of government and non-government organizations have installed GIS with their own finance and are operated by local expert[11]. The Bangladesh Climate Change Strategy Action Plan (BCCSAP) is being implemented in the country since its adoption in 2009 by the Government of Bangladesh. It has been observed that the plan needs to give special concentration to the southwest part given its vulnerability to natural calamities.

## 2.1 COMPLEX ANALYSIS OF GIS

Many different types of spatial analysis can be expressed as database queries in the query language using the spatial abstract data types [1]. The results of a query can be presented in the standard tabular form or on a map display using the Query Shapes to depict the retrieved objects. A spatial analysis can be performed by creating queries which use the operators of the spatial Abstract data types [7]. For example, assume we have created the data model in Figure.1a and b. The first query (Figure.1c) finds all towns with an area greater then 10,000 square meters by applying the Area2Pgn function to the attribute region. The second query (Figure.1d), converts the region to a bounding box and tests if it has overlap (operator &&) with a given rectangle.

```
a) create town (name=text, population=int4, region=POLYGON2)
b) create road (name=text, construct=date, shape=POLYLINE2)
c) retrieve (town.all) where Area2Pgn(town.region)>10.000
d) retrieve (town.all) where Box2Pgn(town.region) && "(0,0,800,400)": : Box2
e) retrieve (road.all) where Length2Pln(road.shape) <  5.000
   and road.construct > " 1Jan   1990"
f) retrieve (town.all) where MinDist2PgnPln(town.region, road.shape) < 500
and road.name= " A12"
```

Figure. 1: A spatial data model analysis.

Analysis can be performed by using joins based on the spatial attributes. The last query (Fig. 1) retrieves every town that lies within 500 meters of one specific road [8].

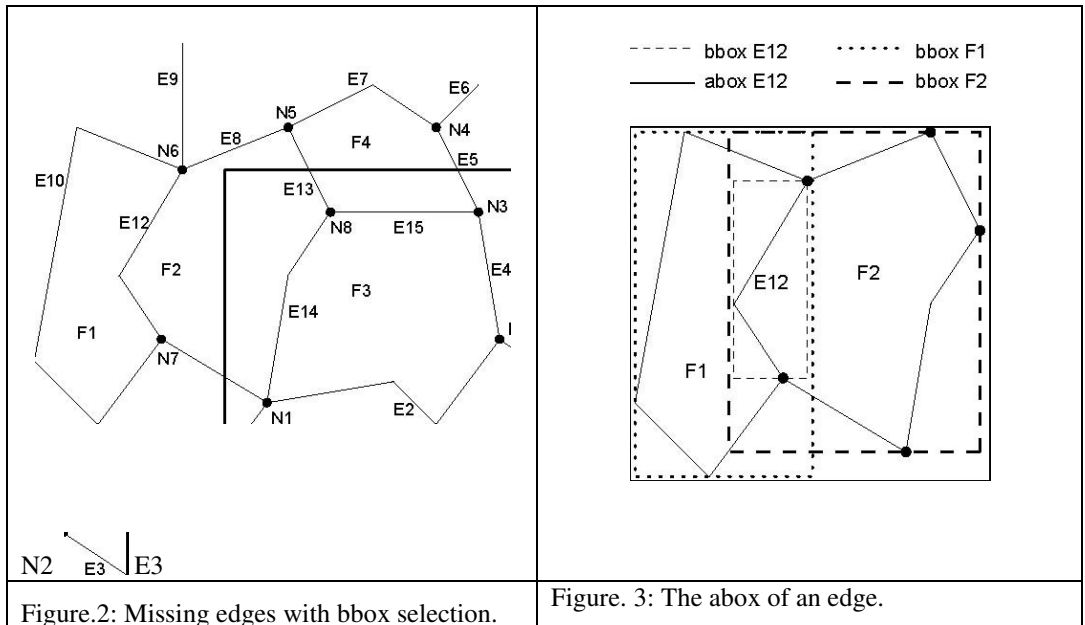

Figure.2: Missing edges with bbox selection.

Figure. 3: The abox of an edge.

When considering the integration of spatial analysis in a GIS two important aspects have to be distinguished: data sharing between the **GIS** and **analysis software**. Several techniques are available for achieving this: ASCII files, binary files, common database, shared memory, dynamic loading of functions into the address space of the core GIS process. Good child classifies the types of integration into the following categories (1992):

1. Stand-alone spatial analysis software and no data exchange:
2. Loose coupling data exchange between GIS and spatial analysis software through ASCII or binary files:
3. Close coupling spatial analysis software using the macro language and 'hooks' of the core GIS: one common data model is used:
4. Full integration of spatial analysis software and the GIS into one program data is in the 'same Program'.

## 2.3 TOPOLOGY

The advantages of using a planar topological structure in a GIS database are well-known: it avoids redundancy when storing common boundaries and it is very suitable for certain types of spatial analysis. The drawback of using a topological structure is that the visualization is often indirect and therefore slower areas in a topologically structured polygonal map. First, the references from areas to edges have to be retrieved. The actual edges have to be retrieved (based on the references) and finally, the polygons have to be reconstructed from the edges. Besides normal polygons we also want to handle polygons with holes or islands. In the list of references to edges, the start of a new island is marked with a 0 separator. In this particular case, besides enabling real-time GIS analysis – anytime, anywhere, the distribution of the GIS information to a wider audience on the Web increase its overall value to the research group. It may also need to mention how Internet was designed and built as a distributed and decentralized system and because its accessibility and actuality, has become a very efficient way to disseminate geospatial data. On the other hand, the last generations of GIS programs provide some enhance features such as interoperability, increased viewing flexibility, more powerful analytical tools and scalability. Finally, the development of Spatial Data Infrastructures (SDIs) based in the Open Geo Spatial Consortium (OGC) specifications have opened promising opportunities for cartographers and GIS experts. The system will integrate with GIS project data in Google Earth also open new opportunities for supplementary visual analysis.

## 3. SAMPLE INPUT VARIABLES

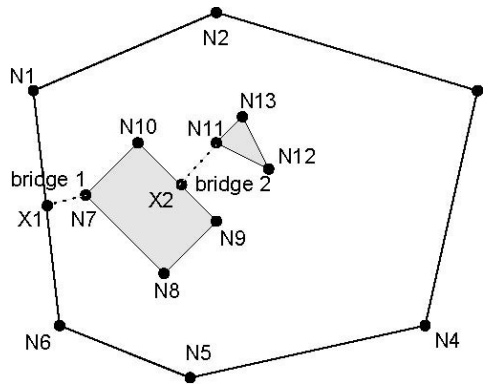

Original polygon: N1, N2, N3, N4, N5, N6
land1: N7, N8, N9, N10 set of input data.
land2: N11, N12, N13 set of input data.

**Algorithm [1]:**

| Figure. 4: A polygonal data model |
|---|
| a. create boundaries (b_id=int4, line=POLYLINE2, bbox=box, abox=box) <br> /* possibly also include a_id_1, a_id_r */ |
| b. create areas (a_id=int4, b_ids=int4[], bbox=box) |
| c. create geo_topol (ind_relname_text, ind_relattr=text, <br> topoltype=text,   /* linear, polygonal network, hierarchy,…. */ <br> ref_count=int   /* #objects referred 1,2,……… or variable  (0)*/ <br> ref_relname=text, ref_relid_text, ref_relvis=text, ref_relbbox=text) |
| d. append geo_topol (ind_relname="areas", ind_relattr="b_ids", <br> topoltype="ind   polygon to polyline", <br> ref_count=0, <br> ref_relname="boundaries",  ref_relid="b_id", <br> ref_relvis="line",  ref_relbbox="abox") |
| e. append geo_dyninfo (relname="boundaries", relattr="line", <br> bboxattr="bbox", dynfunc="bin_pln2_shape", is_bin="t", <br> dynfile="$GEOHOME/dynamic/Geo_Shapes_o" the_oid="oid") |
| f. append geo_dyninfo  (relname="areas", relattr="b_ids", <br> bboxattr="bbox", dynfunc="bin_tpgn2_shape", is_bin="t", <br> dynfile="$GEOHOME/dynamic/Topol2Shapes.o", the_oid="oid", <br> cleanup_func="tpgn2_clean") |

International Journal of Information Sciences and Techniques (IJIST) Vol.3, No.4, July 2013

## 3.1 CCM ANALYSIS

The cross country movement (CCM) problem is also known as the Weighted Region (Least Cost Path) in vector CCM analysis The quality of the returned shortest paths, depends on the fines of the grid, and on the number of move directions permitted. The ideal solution is only found when both tend to infinity .Clearly; this is a very unpractical solution .Therefore, the vector approach, which produces an exact solution[12]. In Raster CCM Analysis Several raster-based algorithms for finding the shortest path have been developed such as the Delphi Method and the Nominal Group Technique [2], have been developed to assign the results. Results are calculated by several geographic variables such as soil type, obstacles, vegetation and overgrowth. Other factors that influence the results are weather conditions and vehicles movement like bus, car and bike. Traversing cost of a given region polygon is uniform within the region but varies between regions and is based on soil, vegetation, etc. Finding an optimal path locating a corridor from a given source location to a given destination that can be used for traveling but also for planning highways, railways, pipelines and other transport systems. The cost function is based on optimization criteria such as time safety, fuel usage, impact, length, etc[1]. There exist raster and vector-based algorithms for this problem. The vector-based approach of Mitchell and Papadimitriou (1991) has been implemented. The first step in the vector algorithm is to apply a constrained Delaunay triangulation (Chew, 1987: Lee & Schachter, 1980) to the polygonal map data. Then a wave-front propagation technique is used to trace the optimal paths from the source.

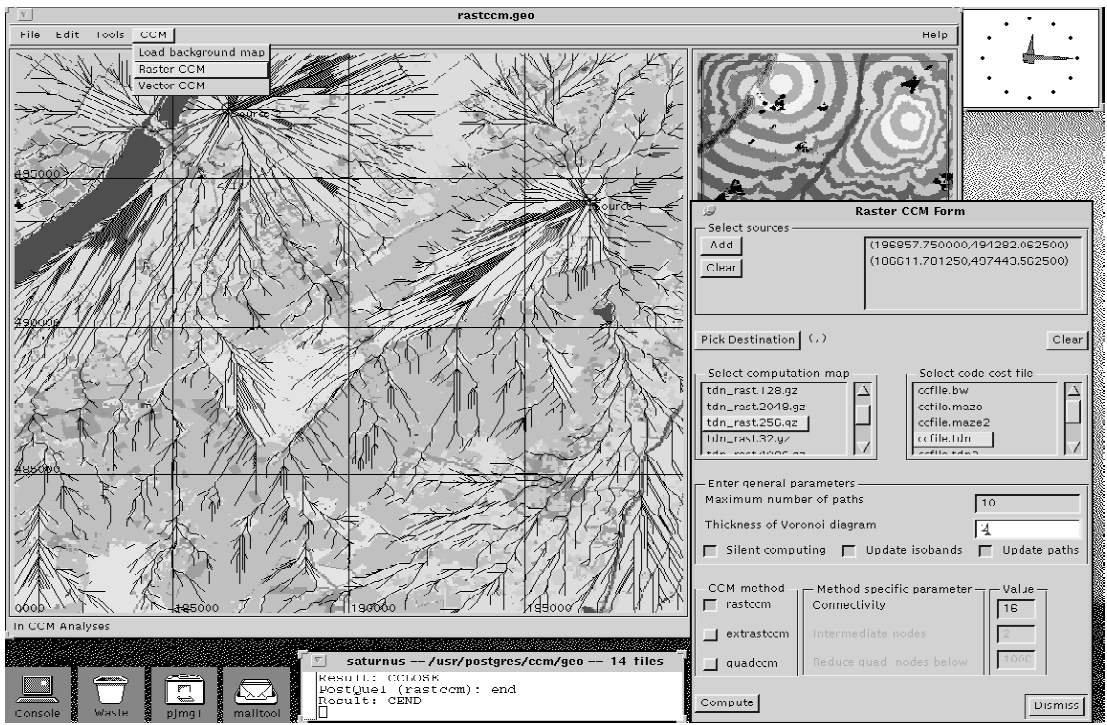

Figure5: User interface of CCM analysis programs display [6].

## 4. APPLICATION AREAS OF GIS

GIS tools are designed for computer aided field data collection which are now used extensively in government, business, and research for a wide range of applications including Facilities Management, Environment and Natural Resources Management, Street Network , Planning and Engineering , Land Information System , and Archaeological analysis [10].

**Facilities Management**: Large scale and precise maps and network investigation are used mainly for utility management. AM/FM is frequently used in large area. Application of GIS in this area have included locating underground pipes and cables, balancing loads in electrical networks, planning facility maintenance, tracking energy use.

**Environment and natural resources management:** Medium or small scale maps and overlay techniques in combination with aerial photographs and satellite images are used for management of natural resources and environmental impact analysis that includes management of wildlife habitat, wild and scenic rivers, recreation resources, floodplains, wetlands, agricultural lands, aquifers, forests.

**Land Information System:** It is used in areas like splitting and subdivision planning, land acquirement, environmental impact policy, water quality management, maintenance of ownership.

**Street-networks**.: GIS has been found to be particularly useful in address matching, location analysis or site selection, development of evacuation plans.

**Planning and Engineering:** Large or medium scale maps and engineering models are used mainly in civil engineering.

**Archaeological analysis:** GIS has been applied to perform complex analysis in archaeology to distinguish history, position, activity and roles.

### 4.1 MAJOR GIS BASED SOFTWARE IN BANGLADESH

ArcInfo and ArcView GIS are common and popular software in the country. ArcView extension tools are being used for advance GIS modelling like Spatial Analyst, 3D Analyst, Network Analyst, Image Analyst, Internet Map Server. The other useable software are ERDAS, ERDAS IMAGINE, IDRISI, Tosca, ER Mapper, SPANS, MapInfo, MapBasic, Imagine, Earth View, Surfer, Lantastic Network, AutoCAD, ArcFM, ArcMap, Map Objects, Arc Objects and ArcGIS[11]. All these softwares can run following operating systems platform:  DOS, Windows and Unix.

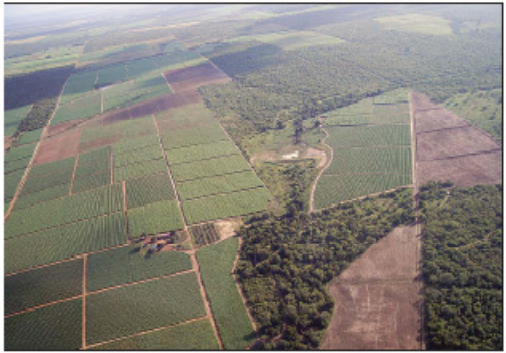 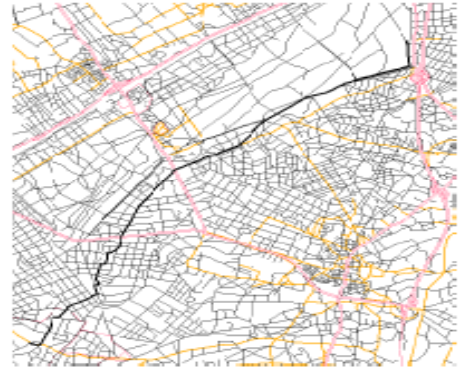

Figure.6: Plain fields                      Figure. 7: Digitized fields

Digitized fields are stratified based on the probability of finding a crop. The core strata used are high, medium, and low cultivation. High, medium, and low refer to the densities of fields within any given area as well as the presence of pivot irrigation and small-scale farming[13]. Stratification is done to increase sampling efficiency. More section points are used in strata where there is a higher possibility of finding crops of interest. This will obtain the most useful data within budget more complicated types of analysis can be performed by using joins based on the spatial attributes. Bangladesh probably has a space orbital path that is currently rented out to Malaysia where a satellite is now being deployed. Soon or later the rental term will expire, then Bangladesh might take actions to set up its own satellite on that orbital to use for GIS purposes as well as for other telecommunication purposes: civil and military. With proper GIS based coordination installed at the national level we would have the advantage to monitor our landscapes for any changes directly from our satellite images. This would allow our government to take necessary measures against any illegal activities in the country's geographical boundaries. Today we have many GIS professionals in the country who are working in different organizations and supporting the academic institutions to develop the field. For example, University of Dhaka, University of Jahangir Nagar, University of Rajshahi, University of Chittagong, the Engineering Staff College of the Institution of Engineers, Bangladesh, Space Research and Remote Sensing Organization (*SPARSO*) and much other organization are offering preparation and academic agenda in the country [10]. Therefore, there is no shortage of experts and professionals in the country to undertake and implement a GIS project at this level.

## 5. GIS ACTIVITY AND IMPLEMENTATION

a. In 1993, the Rajdhani Unnayan Kartipakkha (Rajuk) set up GIS in the Bangladesh. In this organization the main field of GIS application is urban planning. At this point, the GIS action mainly concentrates on mapping and stored data management for development planning of Dhaka Metropolitan Area. Rajuk also prepared urban land use planning map and infrastructure map at strategic 1:50,000 to detail 1:3,960 using spatial and attribute data.

b. In 1995 Roads and Highways Department (RHD) under an Institutional Development Components (IDC) Project sponsored by Overseas Development Agency (ODA) completed GIS mapping programmers to create national transport network. In 1996 the project also successfully built a comprehensive geographical database for the road and rail sectors, which started operation in 1997[14].

4International Journal of Information Sciences and Techniques (IJIST) Vol.3, No.4, July 2013c. GIS technology installed by Survey of Bangladesh (SOB) mainly for making and publishing digital maps. For this reason SOB works in cooperation with other national organizations like SPARRSO, BBS, DLRS and international organizations like JICA (Japan International Cooperation Agency), IGN of France, ITC of Netherlands and Ordinance Survey of England [14].

d. To facilitate remote sensing the major GIS technology installed in 1991 for activities of SPARRSO and other attributes data for various applications in environment and resource fields. The successful projects which are conducted by SPARRSO in this regard are Land use Mapping, Crop estimation, Forest wrap Mapping, Shrimp Culture Potentiality Mapping, Census Mapping and Monitoring of Ecological Changes.

e. Soil Resources Development Institute (SRDI) render supports for preparing Thana Land and Soil utilization Guides including a soil database, soil fertility and land use monitoring, salinity monitoring and preparation of soil and land use related maps[14]. The entire activities of mapping and monitoring systems are GIS allied.

f. Surface Water Modeling Centre (SWMC) is using GIS as a data processing, modeling and planning tool. By using GIS, SWMC is succeeded in monitoring optimum operation of Karnafuli Hydro Power Station, arsenic contamination of groundwater and crop damage assessment [11]. They are also successful in GIS based software development. Interactive Information System (IIS) is one of the key development software, which combines topographic maps prepared under a Geographical Information System and field information of channels, structures, roads, embankments, homesteads stored in a Rational Database Management System (RDMS) [14].

g. The Water Resources Planning Organization (WARPO) prepared and updated 'National Water Resources Database' (NWRD) for preparing the National Water Policy adopted by the Government of Bangladesh[14]. GIS based graphical user interfaces in the front-end and in the back-end database is designed with SQL (Structured Query Language) . The primary activity of NWRD is to meet the demand of water resource planners for a consolidated and reliable data bank[14].

In order to create skilled manpower for the country, most of the universities of Bangladesh installed GIS for their academic programs. The department of Geography and Environment, Jahangirnagar University set up GIS lab in 1992. The following year several other university departments established GIS lab. These are: the department of Geography and Environment, University of Dhaka; the department of Geography and Environmental Studies, Rajshahi University; Urban and Rural Planning Discipline, Khulna University; the Department of Urban and Regional Planning (URP) and Bangladesh University of Engineering and Technology (BUET)[11]. Recently the department of Geology and the department of Soil Science, Water and Environment, from Dhaka University and the department of Geography of Chittagong University also established GIS research lab. Other GIS based installed organizations and companies are: Bangladesh Water Development Board (BWBD), Bangladesh Inland Water Transport Authority (BIWTA), Directorate of Land Records and Surveys (DLRS), CIPROCO Computers Ltd, Cooperation of American Relief in Everywhere (CARE), Danish International Development Agency (DANIDA), Geographical Solutions Research Centre (GSRC) Ltd, Development Design

27

Consultants (DDC), Natural Resources Programs (NRP), Department of Environment (DOE), GEOSERV Ltd, International Centre for Diarrhea Disease Research, Bangladesh (ICDDR,B), Japan International Cooperation Agency (JICA), Banglapedia Project of the Asiatic Society of Bangladesh, The Mappa, etc. The following table represents some common fields and activities of GIS technique with concerned organizations [11]:

| Field of applications | Activities | Organisations |
|---|---|---|
| Agriculture | Monitoring, evaluation and management | BARC, SRDI, MOA |
| Environment | observing, management and modelling for land dreadful conditions; climate and weather modelling, forecasting and prediction; river and coastal corrosion modelling and flood management | SPARRSO, EGIS, SWMC, DOE, MOA, CARE |
| Health | Areal division of dissimilar diseases in relation to environmental issues. | ICDDR,B; DPHE |
| Forestry | Planning, management, map arrange for position specific matching. | DOF |
| Regional/Local planning | Infrastructure development programme, Land Registration Development of plans, preservation, management. | Rajuk, DLRS, SPARRSO, LGED, CARE |
| Research and education | Various locations problem solution from personal to national level. | Educational institutions and Consultant Organisations |
| Resource | Management, planning, monitoring, recording | SPARRSO, DOF, BCAS, EGIS, LGED |
| Social studies | Demographic trends and developments analysis | BBS, Educational institutions |
| Transport network | Planning and management | RHD, LGED, SOB |
| Others | Site and Location Information, Services, Thematic mapping, Topographical mapping, Consultancy etc. | SOB, LGED, DLRS, WARPO, Banglapedia, different companies |

## 6. RESULTS AND DISCUSSION

This paper describes about the GIS, Digital Mapping, Internet concepts and tools that can successfully contribute in the modeling, analysis and image representation according to find out the solutions of various problems like soil fertility, proper land use, land dreadful conditions estimation and possible control procedures of land resources. This database constitutes the groundwork feedback for the agricultural production by using object detection, object tracking and field mapping. Through the studied GIS mapping of the agricultural land resources statistics can be made of how much land is suitable for harvest and each year how much land we are losing. Study shows the mapping of the cultivation land will give us idea of how much crop we have to import from other countries also. The system will give assistance and give transparency to the food stock of the country according to the harvest. For this research the new interactive structure system can successfully integrate with the GIS project data. The study first finds highly accurate cluster images and partial images, obtain user feedback to merge or correct these digests, and then after considering various factors the additional visual scrutiny complete the data partitioning.

## 7. CONCLUSION

In this paper we have described an assessment for storing and visualizing topological data in a GIS for cultivable land by matching the pattern through direct analysis functions. Thus these assessment can be used for planning field campaigns such as mapping of foliations relative to the topography near potential environment allows the integration of all aspects of a complex GIS system to be integrate directly of third parties in a coherent interpreted frame work. They can reorganize a collection of data into a different location .We feel further research directed towards cooperation between on-line GIS mapping application and the end-user will prove fruitful. It also enables the implementation of tightly coupled analysis modules with respect to the data and the Graphical User Interface. This is beneficial for the performance because data transfers are condensed the acknowledgements and approach the techniques to reduce the number of search nodes looks very promising.

**AUTHORS**

**Yeasir Fathah Rumi**
Assistant Professor & department coordinator, of Computer Science & Engineering department, Bangladesh University, Dhaka, Bangladesh. He completed B.Sc. Engineering degree in Computer Science & Engineering from The University of Asia pacific and later on M.Sc. in Computer Science. He also studied in Edith Cowan University, Australia. After completing M.Sc. in CSE, he joined Bangladesh University and currently working as an Assistant Professor. He has successfully completed Cisco Certified Network Associate(CCNA) and Microsoft Certified IT Professional (MCITP) on Server 2008 platform, He successfully trained in GTI. He has nominated Book reviewer of National Curriculum of Textbook Board, Dhaka. 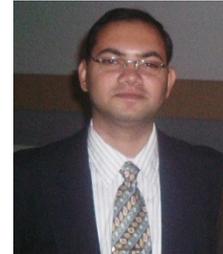
His research interest includes Microprocessor and assembly language ,Operating System, Networking, VLSI Design, Digital Logic, Peripheral & Interfacing, Digital System Design and Web Programming. He has many international and national research publications and conference paper. His email address: yeasirfathah@yahoo.com

**Uzzal Kumar Prodhan**
Assistant Professor, Department of Computer Science & Engineering, Jatiya Kabi Kazi Nazrul Islam University, Trishal, Mymensingh, Bangladesh. He has completed his M.Sc. and B.Sc. from the department of Computer Science & Engineering, Islamic University, Bangladesh. After completing M.Sc in CSE, he joined Bangladesh University as a Lecturer & joined in Jatia Kabi Kazi Nazrul Islam University as an Assistant Professor. In his long teaching life he was appointed as a head examiner in Computer Technology by Bangladesh Technical Education Board, Dhaka. Due to his teaching interest he was selected as a Book reviewer of National Curriculum of Textbook Board, Dhaka. He has successfully completed Microsoft 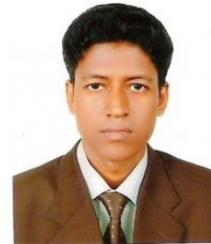
Certified IT Professional (MCITP) on Server 2008 platform. His research interest includes Artificial Intelligence, Neural Network, Cryptography, Computer Architecture and Organization and Pattern Recognition

**Mohammed Ibrahim Hussain**
Assistant Professor & semester program coordinator, Department of Computer Science & Engineering, Bangladesh University, Dhaka, Bangladesh. He had M.Sc. in E-Commerce from London, UK. He got B.Sc degree in Computer Science & Engineering, Kiev, Ukraine. After completing M.Sc in CSE, he joined Bangladesh University as a Lecturer. He has successfully completed Cisco Certified Network Associate(CCNA) and Microsoft Certified IT Professional (MCITP) on Server 2008 platform.He is also a nominated Book reviewer of National Curriculum of Textbook Board, Dhaka. His research interest includes Operating System, Networking, VLSI 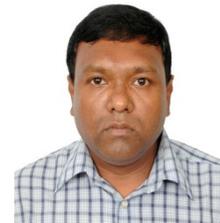
Design,Digital, Logic,Perpheral Interfacing,Digital System Design and Web Programming. He has many international and national research publications. His email addresses are ibrahim.hussain@bu.edu.bd, ibrahim180772@gmail.com .

**A.H.M Shahariar Parvez**
Assistant Professor, Department of Computer Science & Engineering, Bangladesh University, Bangladesh. He has completed his M.Sc. in Computer Science & Engineering, Kiev, Ukraine. He also got another M.Sc. in E-Commerce from UK. He got B.Sc degree in Computer Science & Engineering, Kiev, Ukraine. After completing M.Sc in CSE, he joined Bangladesh University as a Lecturer. He has successfully completed Microsoft Certified IT Professional (MCITP) on Server 2008 platform. His research interest includes Operating System, Digital Logic, Digital System Design and Electrical Drivers and Instrument. He has many international and national research publications. His email addresses are shariar.parvez@bu.edu.bd, sha0131@yahoo.com.

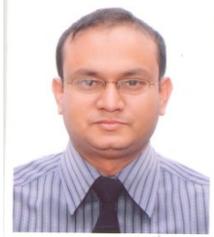

**Md. Ali Hossain**
Md. Ali Hossain was born in Manikganj, Bangladesh. He received the B.Sc. and M.Sc. degrees from the Department of Computer Science and Engineering, University of Islamic University, Kushtia, Bangladesh, in 2008 and 2009, respectively. He is serving as a Lecturer with the Department of Computer Science and Engineering (CSE), Bangladesh University, Dhaka.  His current research interests include Biomedical Imaging, Biomedical Signal, Speech Processing Bioinformatics and Cryptography. Mr. Ali Hossain  is an Associate Member of the Bangladesh Computer Society and Executive Member of  Islamic University Computer Association (IUCA). He has many international research publications. His email address is ali.cse.bd@gmail.com.

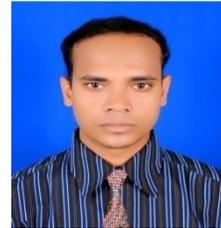